\documentclass[aps,prl,10pt,twocolumn,preprintnumbers,superscriptaddress,showpacs,floatfix,nofootinbib]{revtex4-2}
\usepackage{textcomp,gensymb}
\usepackage{hyperref}
\usepackage{graphicx}
\usepackage{amsfonts,amsmath,amssymb,bm,bbm}
\usepackage{color,soul}
\usepackage[dvipsnames]{xcolor}
\usepackage{slashed}
\usepackage{flushend}
\usepackage{physics}
\usepackage{graphicx}
\usepackage{dcolumn}
\usepackage{float}
\usepackage{ragged2e}
\usepackage{subcaption}

\hypersetup{
    pdfnewwindow=true,      
    colorlinks=true,       
    linkcolor=blue,          
    citecolor=blue,        
    filecolor=blue,      
    urlcolor=blue        
}

\begin{document}
\preprint{MI-HET-859, MITP-25-043}
\title{`Dark' Matter Effect as a Novel Solution to the KM3-230213A Puzzle}

\author{P. S. Bhupal Dev}
\email{bdev@wustl.edu}
\affiliation{Department of Physics and McDonnell Center for the Space Sciences, Washington University, St. Louis, MO 63130, USA}
\affiliation{PRISMA$^+$ Cluster of Excellence \& Mainz Institute for Theoretical Physics, 
Johannes Gutenberg-Universit\"{a}t Mainz, 55099 Mainz, Germany}

\author{Bhaskar Dutta}
\email{dutta@tamu.edu}
\affiliation{Mitchell Institute for Fundamental Physics and Astronomy, Department of Physics and Astronomy, Texas A\&M University, College Station, TX 77843, USA}

\author{Aparajitha Karthikeyan}
\email{aparajitha_96@tamu.edu}
\affiliation{Mitchell Institute for Fundamental Physics and Astronomy, Department of Physics and Astronomy, Texas A\&M University, College Station, TX 77843, USA}

\author{Writasree Maitra}
\email{m.writasree@wustl.edu}
\affiliation{Department of Physics, Washington University, St. Louis, MO 63130, USA}

\author{Louis E. Strigari}
\email{strigari@tamu.edu}
\affiliation{Mitchell Institute for Fundamental Physics and Astronomy, Department of Physics and Astronomy, Texas A\&M University, College Station, TX 77843, USA}

\author{Ankur Verma}
\email{averma1@tamu.edu}
\affiliation{Mitchell Institute for Fundamental Physics and Astronomy, Department of Physics and Astronomy, Texas A\&M University, College Station, TX 77843, USA}
\begin{abstract}
The recent KM3NeT observation of an ${\cal O}(100~{\rm PeV})$ event KM3-230213A is puzzling because IceCube  with much larger effective area times exposure has not found any such events. 
We propose a novel solution to this conundrum in terms of dark matter (DM) scattering in the Earth's crust. 
We show that intermediate dark-sector particles that decay into muons are copiously produced when high-energy ($\sim100~\text{PeV}$) DM propagates through a sufficient amount of Earth overburden. The same interactions responsible for DM scattering in Earth also source the boosted DM flux from a high-luminosity blazar. We address the non-observation of similar events at IceCube via two examples of weakly coupled long-lived dark sector scenarios that satisfy all existing constraints. We calculate the corresponding dark sector cross sections, lifetimes and blazar luminosities required to yield one event at KM3NeT, and also predict the number of IceCube events for these parameters that can be tested very soon. Our proposed DM explanation of the event can also be distinguished from a neutrino-induced event  in future high-energy neutrino flavor analyses, large-scale DM direct detection experiments, as well as at future colliders. 
\end{abstract}
\maketitle
{\textbf {Introduction.--}} The KM3NeT collaboration has recently reported the detection of an ultra-high-energy throughgoing muon event with energy $120^{+110}_{-60}$ PeV~\cite{KM3NeT:2025npi}. This is the highest energy event ever detected by a neutrino telescope, surpassing the previous record set by IceCube~\cite{IceCube:2023agq} by almost an order of magnitude. Since this is a throughgoing muon, the parent particle, assumed to be a neutrino in the KM3NeT analysis, must carry even higher energy, estimated to be in the range of 110--790 PeV with a median energy of 220 PeV. The excellent angular resolution for muon tracks enabled KM3NeT to reconstruct the direction of the event to be near-horizontal, originating $0.6^\circ$ above the horizon at an azimuth of $259.8^\circ$ with an uncertainty of $1.5^\circ$ at 68\% confidence level (CL). In equatorial coordinates (J2000), this event points to the Southern hemisphere with right ascension (RA) of 94.3$^\circ$ and declination angle (dec.) of $-7.8^\circ$.        

Two major issues make this event rather unusual: (i) Why did this event evade detection at IceCube, which has 10 times more exposure and about 20 times larger effective area~\cite{IceCube:2014vjc} than the current KM3NeT? The event is located about $8^\circ$ above the horizon for IceCube; in this direction, IceCube has the maximum effective area. It is true that the event will be downgoing for IceCube and it could be confused with a cosmic-ray induced event, although given the enormous energy, that seems unlikely. The observed tension between KM3NeT and other datasets, including  null observations above tens of PeV from the IceCube and Pierre
Auger observatories, is at the level of 2.5--3.6$\sigma$~\cite{Li:2025tqf, KM3NeT:2025ccp}. To beat IceCube's advantage of exposure time, a transient point source explanation~\cite{KM3NeT:2025bxl} seems more plausible than a diffuse cosmogenic~\cite{KM3NeT:2025vut} or galactic~\cite{KM3NeT:2025aps} source. However, this leads to another question: (ii) What kind of cosmic accelerators can produce such a high-energy particle? Given the enormous energy of the event, the source is most likely extragalactic. Blazars are among the most powerful cosmic accelerators which are promising neutrino sources as well, as confirmed by the  multi-messenger observation of the TXS 0506+056 event~\cite{IceCube:2018cha}. In fact, KM3NeT has identified 17 blazars within $3^\circ$ of the KM3-230213A location in the sky through their multiwavelength properties~\cite{KM3NeT:2025bxl}. Taking a typical redshift of $z\approx 1$ for these sources, the estimated source luminosity for the blazar jet from the inferred neutrino luminosity to explain the event is $L_p\simeq 10^{50}$ erg/s~\cite{KM3NeT:2025bxl}, orders of magnitude larger than a typical blazar luminosity of $10^{45}$ erg/s without beaming. Even with a beaming factor of $10^3$, the blazar needs to be flaring for ${\cal O}(100)$ years to meet the required neutrino flux, thus putting the standard interpretation again in tension with IceCube.     

Recently, there have been several  attempts at understanding the origin of the KM3-230213A event in terms of beyond-the-Standard Model (BSM) physics, such as decaying heavy dark matter (DM)~\cite{Borah:2025igh, Kohri:2025bsn, Narita:2025udw,Jho:2025gaf, Barman:2025hoz,Murase:2025uwv, Khan:2025gxs}, primordial black hole evaporation~\cite{Boccia:2025hpm,Jiang:2025blz,Klipfel:2025jql,Dvali:2025ktz, Choi:2025hqt}, Lorentz invariance violation~\cite{Satunin:2025uui, KM3NeT:2025mfl,Cattaneo:2025uxk, Yang:2025kfr, Wang:2025lgn},   neutrino non-standard interactions (NSI)~\cite{Brdar:2025azm, He:2025bex}, etc. However, none of these BSM explanations address the two above-mentioned issues. Only Ref.~\cite{Brdar:2025azm} addresses the tension with IceCube using non-standard neutrino matter effect. Here we make the first ambitious attempt to simultaneously address both (i) and (ii) in a self-consistent way. 

\begin{figure*}[t!]
  \centering
  \begin{tabular}{
     @{} >{\centering\arraybackslash}m{0.33\linewidth}
     >{\centering\arraybackslash}m{0.33\linewidth}
     >{\centering\arraybackslash}m{0.33\linewidth} @{} }
    \hline
    \textbf{Blazar Production ($p\gamma$)} &
    \textbf{Scattering in Earth}   &
    \textbf{Decay $\to\mu^+\mu^-$} \\
    \hline

    \subcaptionbox{}[\linewidth]{\includegraphics[]{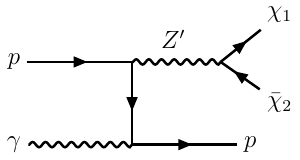}
      } &

    \subcaptionbox{}[\linewidth]{\includegraphics[]{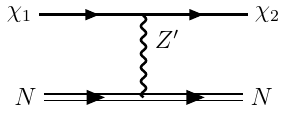}
    
    } &

    \subcaptionbox{}[\linewidth]{\includegraphics[]{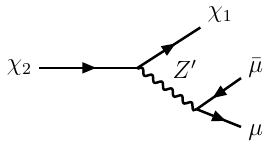}} \\
            \hline
    \subcaptionbox{}[\linewidth]{\includegraphics[]{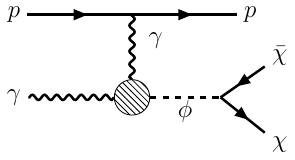}} &

    \subcaptionbox{}[\linewidth]{\includegraphics[]{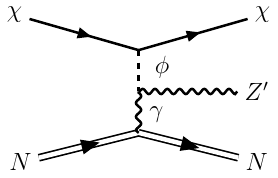}} &
    
    \subcaptionbox{}[\linewidth]{\includegraphics[]{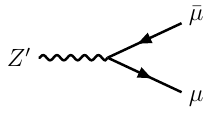}} \\
    \hline
  \end{tabular}
\captionsetup{justification=Justified}
  \caption{Signal chain for blazar-produced DM arriving at KM3NeT.
           \emph{Upper row}: Inelastic DM model with a vector mediator~$Z'$.
           \emph{Lower row}: $2\to 3$ scattering with a scalar mediator~$\phi$.
           Columns show, from left to right, DM production in
           $p\gamma$ collisions at blazars, scattering on terrestrial
           nuclei, and the mediator decay that yields the observable
           $\mu^+\mu^-$ pair.}
  \label{fig:combined_feynman}
\end{figure*}

To this end, we propose that the KM3-230213A event is {\it not} caused by a neutrino, but by a DM. It cannot be a diffuse source of DM though, like the decaying DM solution in  Refs.~\cite{Borah:2025igh, Kohri:2025bsn, Narita:2025udw,Jho:2025gaf, Barman:2025hoz,Murase:2025uwv, Khan:2025gxs}, which is ruled out by the gamma-ray and neutrino constraints (see Supplemental Section I). Instead, we consider a transient source of boosted DM that scatters in the Earth matter -- dubbed as the `dark' matter effect.  For concreteness, we assume a fermion DM scattering via a vector/scalar mediator. We entertain two solutions here: (i) $2\to 2$ up-scattering of an inelastic DM via a vector mediator, $\chi_1 N\to \chi_2 N$ (where $N$ stands for nucleons, i.e. protons and neutrons), followed by the de-excitation of the heavier state $\chi_2\to \chi_1 \mu^+ \mu^-$; and (ii) $2\to 3$ DM scattering via a scalar mediator, $\chi N\to \chi N Z'$, followed by $Z' \to \mu^+\mu^-$; see Fig.~\ref{fig:combined_feynman}. Note that it is currently impossible for KM3NeT (or IceCube) to distinguish a single muon from a highly collimated muon pair just using the stochastic energy loss information. Another novelty of our solution is that the same interactions responsible for DM scattering on Earth could also produce the DM and boost it to ${\cal O}$(100) PeV energy via $p\gamma$ processes in a cosmic-ray accelerator environment, like blazars.\footnote{Blazar-boosted DM has been studied in other contexts, but using ambient DM halos often involving a spike profile~\cite{Gorchtein:2010xa, Wang:2021jic, Granelli:2022ysi, DeMarchi:2024riu, Wang:2025ztb}.} The highly boosted DM in our case is assumed to come from a extragalactic transient point source in the Southern sky, most likely a flaring blazar~\cite{KM3NeT:2025bxl,Filipovic:2025ulm,  Dzhatdoev:2025sdi, Neronov:2025jfj, Podlesnyi:2025aqb}, but the details of the source are not so much relevant for our analysis, as long as the DM production rate is comparable to or higher than the neutrino production rate, which can be easily ensured at such high energies with suitable choice of parameters (see Supplemental Section II). 

The crux of our solution is that when sufficiently energetic DM enters the Earth, it can efficiently upscatter to produce an intermediate dark sector particle, transferring almost all its energy to it, which subsequently decays into muons after traversing some overburden distance. Additionally, the near-horizontal source direction for KM3NeT is crucial for this solution to work. For IceCube located at the South Pole, the source coordinates point to $8^\circ$ above the horizon. Analogous to the neutrino matter effect scenario in Ref.~\cite{Brdar:2025azm}, the DM produced in the source will be downgoing for IceCube, and will encounter much less Earth overburden (about 14 km) compared to KM3NeT (about 147 km). We further realize that the $1.5^0$ uncertainty in the source location results in potentially larger overburden distances for KM3NeT ($59- 418$~km) than for IceCube ($12-17$~km).
Therefore, the DM flux from the blazar has a larger probability to upscatter inside Earth's crust and produce muon events at KM3NeT than at IceCube. 

{\textbf{The Models.--}} We consider two DM scenarios to demonstrate our concept and explain the KM3NeT event. The first one involves a two-component inelastic DM $(\chi_1$, $\chi_2)$ with masses $m_{\chi_1,\chi_2}$ that couples to a vector boson $Z'$ with mass $m_{Z'}$ (see e.g.,~Refs.~\cite{Tucker-Smith:2001myb,Izaguirre:2014dua,Giudice:2017zke,Dutta:2019fxn}). Here, $Z'$ behaves as a portal between the DM and Standard Model (SM) sectors. The relevant interaction Lagrangian is 
\begin{equation}
    \begin{aligned}
        -\mathcal{L}_{2\to 2} & \supset  m_{\chi_1}\bar{\chi}_1\chi_{1} + m_{\chi_2}\bar{\chi}_2\chi_{2}  \\ 
        & + \frac{1}{2}m_{Z'}^2 Z'^{\alpha}Z'_{\alpha} + \left(g_\chi\bar{\chi}_2\gamma^{\alpha}\chi_{1}Z'_\alpha+{\rm h.c.}\right)\\
        & + Z'_{\alpha}\big(g_{Z'\mu}\bar{\mu}\gamma^{\alpha}\mu   + g_{Z'q}\sum_{q = u,d} \bar{q}\gamma^{\alpha}q \big)  \, ,
    \end{aligned}    
\end{equation}
where $g_\chi$, $g_{Z'f}~(f = q,\mu)$  denote the coupling strengths of $Z'$ with the DM and with the SM fermions, respectively. For our solution to work, the $Z'$ must couple to the first-generation quarks ($u,d$) and to the muon; the couplings to other SM fermions is optional and will come with additional constraints.  

The second scenario involves a single-component DM $\chi$ with mass $m_{\chi}$ that couples to a new scalar $\phi$ with mass $m_\phi$, which interacts with the SM sector through an effective vertex involving a $Z'$ and the SM photon~\cite{deNiverville:2018hrc, Dutta:2025fgz}. The interaction Lagrangian is
\begin{equation}
    \begin{aligned}
        -\mathcal{L}_{2\to 3} &\supset  m_{\chi}\bar{\chi}\chi + \frac{1}{2}m_{Z'}^2Z'^{\alpha}Z'_{\alpha} + \frac{1}{2}m_{\phi}^2\phi^2 + g_{\chi}\bar{\chi}\chi \phi\\ & + \frac{1}{2}g_{\phi Z' \gamma}\phi F^{\alpha \beta}F'_{\alpha \beta} + g_{Z'}Z'_{\alpha}\bar{\mu}\gamma^{\alpha}\mu \, ,
    \end{aligned}   
\end{equation}
where $g_{\phi Z'\gamma}$ is an effective dimension-5 interaction that can be realized at loop-level in a ultraviolet-complete theory, e.g. via a 3rd-generation SM fermion loop coupled to the SM Higgs mixed with a singlet scalar. One could replace the scalar with an axion-like particle~\cite{deNiverville:2018hrc, Deniverville:2020rbv, Hook:2021ous, Zhevlakov:2022vio, Jodlowski:2024lab}, which would lead to the same inferences.

As shown in Fig.~\ref{fig:combined_feynman}, these interaction Lagrangians enable both the production of the DM from $p\gamma$ collisions in a blazar environment, and its  scattering with the Earth matter to yield the observable muon signal at KM3NeT.  

{\textbf {Events from DM Scattering.--}}
 We determine the flux from a blazar with an intrinsic ``DM" luminosity $L_{\chi}$ (in units of erg/s), located at a luminosity distance $d_L \approx 7~\text{Gpc}$  from Earth, corresponding to a source redshift $z\approx 1$. The DM flux from the blazar is boosted into a narrow cone in the direction towards the Earth, resulting in an enhanced flux relative to an isotropic source model.\footnote{This is why a diffuse source of boosted DM, e.g. induced by high-energy cosmic rays~\cite{Bringmann:2018cvk,Ema:2018bih,Cappiello:2018hsu, Alvey:2019zaa,Dent:2019krz,Bell:2023sdq}, does not work for us because the resulting flux is too small at the required energies.} Though the beaming factor is generally model-dependent, following Ref.~\cite{KM3NeT:2025bxl}, we parameterize it by   $f_{\text{beam}}$. For instance, $f_{\rm beam} = 1$ for isotropic emission and $f_{\rm beam} \approx 10^3$ when the emission is concentrated within $4^\circ$ around the jet direction, which is common for electromagnetic emission of blazars as shown by population studies~\cite{Lister:2019ttx}. 
The differential DM flux (in $[\text{GeV$^{-1}$~s$^{-1}$~cm}^{-2}]$) can be written as
\begin{equation}
    \frac{d\phi}{dE_{\chi}} = \frac{L_{\chi} f_{\text{beam}}}{4\pi d_L^2 E_{\chi}^2} \, .
    \label{eq:blazar_beamed}
\end{equation}
Note there are small corrections to this equation depending on the shape of the energy spectrum.

The number of events at a given detector can be calculated as follows:
\begin{equation}
    N_{\rm{evt}} =  T_{\text{exp}}\int_{E_{\chi}^{\rm{min}}}^{E_\chi^{\rm{max}}}dE_{\chi_i} \frac{d\phi}{dE_{\chi_i}}A_{\text{eff}}(N_\text{PMT}, E_{\chi_i},E_{\text{th}}, d_{b}, \delta) \, ,
    \label{eq:nevents_prior}
\end{equation}
where $A_{\text{eff}}$ is the effective area of a detector with energy threshold $E_{\text{th}}$ and dimension $\delta$ along the direction of muon propagation where the Earth's overburden is $d_{b}$. The number of PMTs triggered $N_\text{PMT}$ also dictates the effective area, as it serves as a prior on the energy of the muons produced by the intermediate long-lived particles at the detector. As for the exposure time $T_{\rm exp}$, since KM3NeT has been collecting data for 335 days, we take $T_{\rm{KM3}} \sim 0.9$~yrs. IceCube has been collecting data for the last $\sim 10$ years; so ideally, $T_{\rm{IC}} \sim 10$~yrs. However, for a transient point source like a flaring blazar, the actual exposure depends on the flaring time. Here we assume that the high-luminosity blazar responsible for the KM3-230213A event was actively flaring for $2$ years including the KM3NeT observation window, so $T_{\rm{IC}} \sim 2$~yrs. Using a 1-year flaring window as reported for some potential sources in Ref.~\cite{KM3NeT:2025bxl} would only improve the compatibility of the KM3NeT event with IceCube non-observation, while a larger flaring window would result in more events at IceCube. The important point here is that due to an enhanced cross section and effective area for the DM scattering, compared to the neutrino scattering solution, we can afford to explain the KM3NeT event with a smaller flaring period for a given source luminosity, thus alleviating the tension with IceCube. We discuss the details on effective area for single scattering and multiple scattering scenarios in Supplemental Section III.


\begin{figure*}[t!]
    \centering
    \begin{subfigure}{.47\textwidth}
        \includegraphics[width = \textwidth]{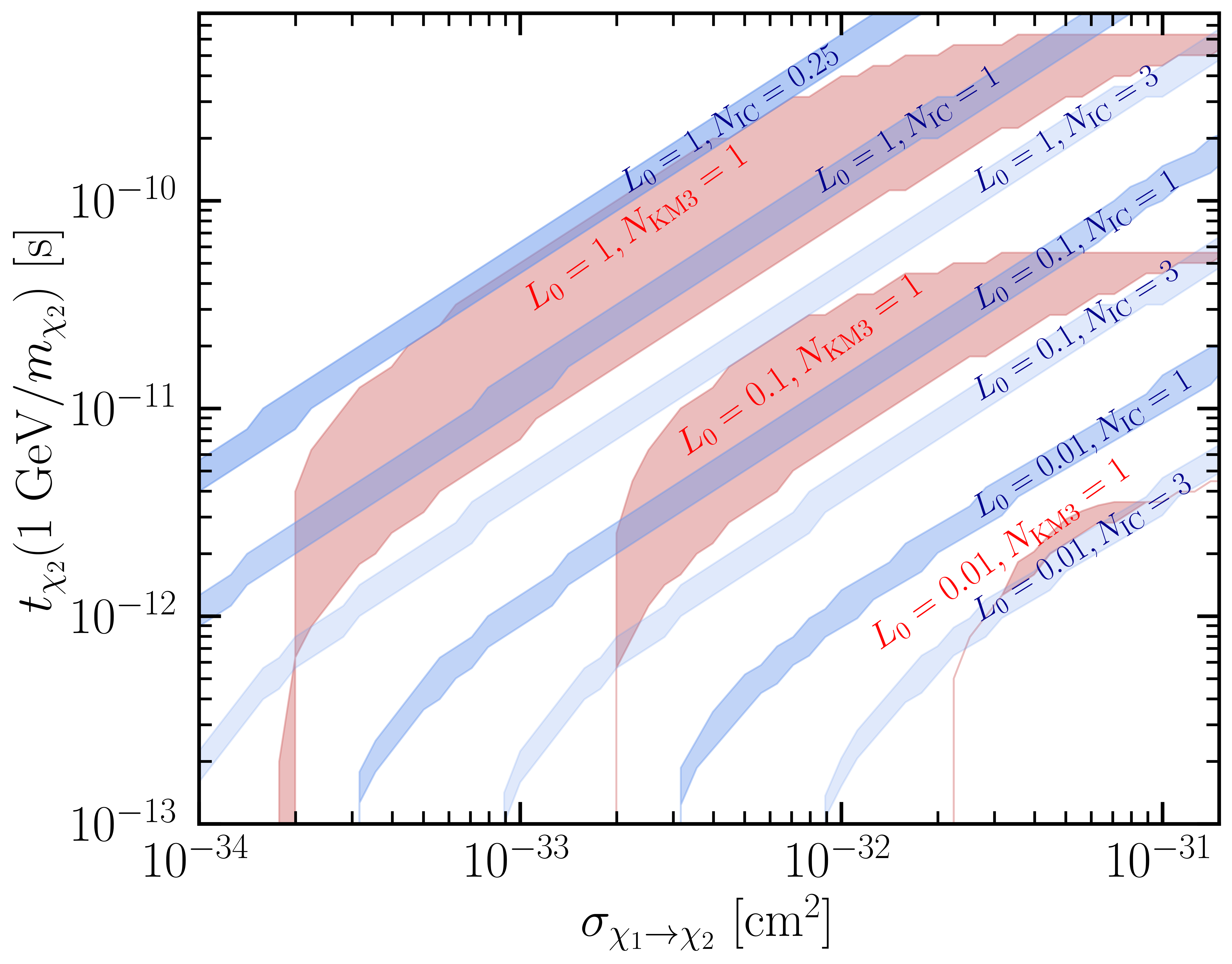}
        \caption{}
        \label{fig:csversust0_inel}
    \end{subfigure}
    \begin{subfigure}{.49\textwidth}
        \includegraphics[width = \textwidth]{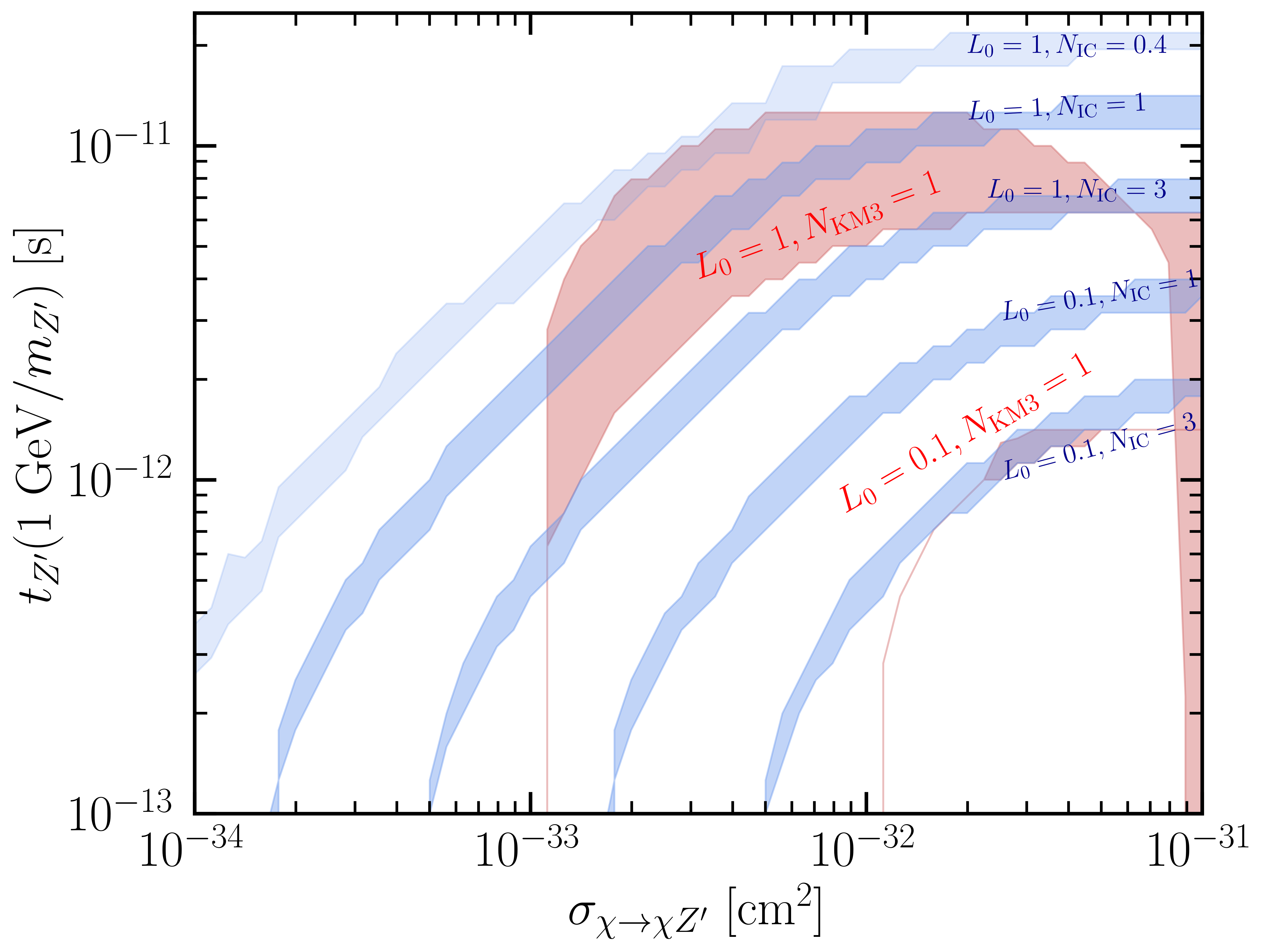}
        \caption{}
        \label{fig:csversust0_el}
    \end{subfigure}
    \captionsetup{justification=Justified}
    \caption{The preferred range of cross sections and rest-frame lifetimes for (a) $2\to 2$ and (b) $2\to 3$ DM scattering scenarios. The peach bands show the parameters corresponding to one event at KM3NeT, and the blue bands show the corresponding events at IceCube (number of events on the band). Various source luminosities are parametrized such that $L_{\chi} = L_0\times 10^{48}~\text{erg/s}$, and $f_{\text{beam}} = 10^3$.}
    \label{fig:csversust0}
\end{figure*}

{\textbf{Sensitivities.--}}
From the sky map in the direction of KM3-230213A~\cite{KM3NeT:2025npi}, we find that the Earth overburden traveled by DM arising from blazars in the $1.5^\circ$ uncertainty region around KM3-230213A is between 59~km and 418~km. The corresponding range of overburden to reach IceCube is between 12~km and 17~km. Since the overburden distance at KM3NeT can be $\sim 35$ times  than at IceCube, KM3NeT is more sensitive to the DM parameter space that has scattering and decay mean free path lengths (MFPLs)  that are of $\mathcal{O}(100~\text{km})$. 

Figure~\ref{fig:csversust0} shows the iso-event contours for KM3NeT, and the corresponding predictions at IceCube, in the space of cross section ($\chi_i \to X$) and lifetime ($X \to \mu^+ \mu^-$)  for various blazar luminosities. The peach bands correspond  to those that give rise to one event at KM3NeT. The blue bands show the parameters that correspond to a particular number of events at IceCube. For example, when $L_{\chi} = 10^{47}~\text{erg/s}$, we find that along the range of parameters that give rise to one event at KM3NeT (corresponding to the uncertainty in the overburden), the number of events at IceCube can vary between 0.3 and 3. Thus, there exists some viable parameter space which explains the non-observation of this event at IceCube. 

One of the salient features in these sensitivites is that the lifetime required for a single event increases as a function of cross section. This is due to the fact that the mean-free path length for scattering is inversely proportional to the cross section. Therefore, the decrease in mean-free path length with an increase in cross section is compensated by increasing the lifetime, or the mean decay length. Another feature we observe is that for larger DM luminosities, the required cross section to produce can be lessened. Since the overburden faced by KM3NeT is larger than IceCube, this allows for more events at KM3NeT for characteristically larger MFPL. For example, this is observed in Fig.~\ref{fig:csversust0_inel} where the cross sections and lifetimes for $10^{46}~\text{erg/s}$ luminosity give rise to $\sim 3$ times more events at IceCube, whereas when the luminosity is $10^{48}~\text{erg/s}$, KM3NeT can observe up to 4 times more events compared to IceCube. 


Since it is more feasible for inelastic DM to undergo multiple scattering than the single-component DM, (when $m_{\chi_2} - m_{\chi_1} = m_{Z'}$), we find that the sensitivity of KM3NeT is much larger in the inelastic DM scenario, as illustrated in Fig.~\ref{fig:csversust0_inel}. KM3NeT's sensitivity to elastic DM in Fig.~\ref{fig:csversust0_el} is constrained mainly due to the energy loss and probability of the mean energy considered. Although the sensitivity can be enhanced by including the total energy spectrum of $Z'$, we still observe that elastic DM produced from a blazar with $L_\chi = 10^{48}~\text{erg/s}$ can produce at least twice as many events at KM3NeT than IceCube.

Note that the blazar luminosities required for our solution to work are consistent with the observed luminosity distributions of a large population of blazars~\cite{Ajello:2013lka}.  We find that the DM solution is more favored than the SM neutrino solution, even if we assume the same DM and neutrino luminosities, due to different effective areas. For larger overburdens, the effective area of DM is enhanced by upscattering into intermediate particles $\chi_2/Z'$. However, the effective area for the SM neutrino is smaller due to the requirement that the neutrinos must scatter close to the detector, because the resulting high-energy muon from the charged-current process loses energy rapidly as it traverses matter, and having a higher-energy neutrino to compensate for this will come with a smaller flux.


\begin{figure*}[t!]
    \centering
    \begin{subfigure}{.48\textwidth}
        \includegraphics[width = \textwidth]{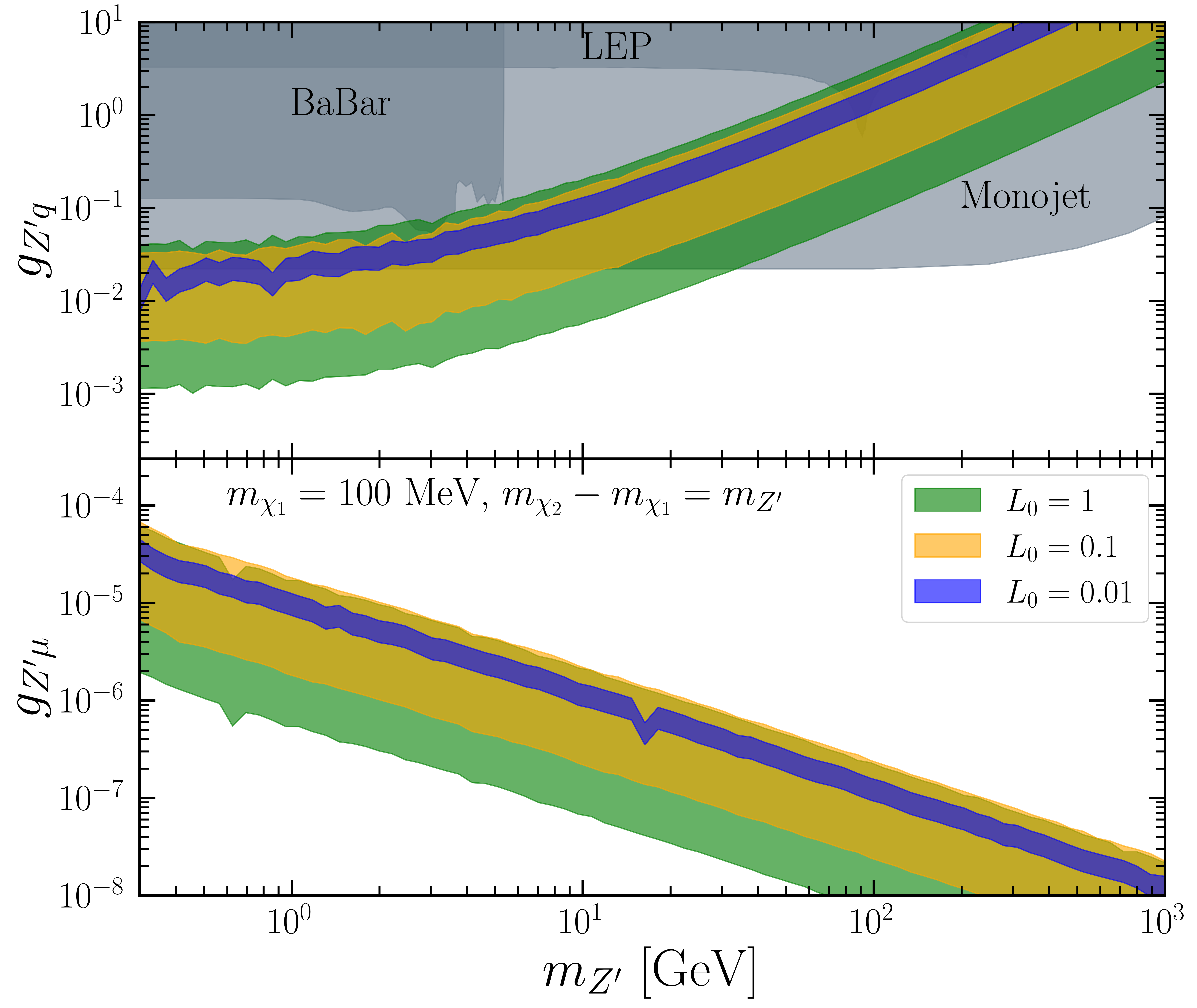}
        \caption{}
        \label{fig:coupmass_inel}
    \end{subfigure}
    \begin{subfigure}{.48\textwidth}
        \includegraphics[width = \textwidth]{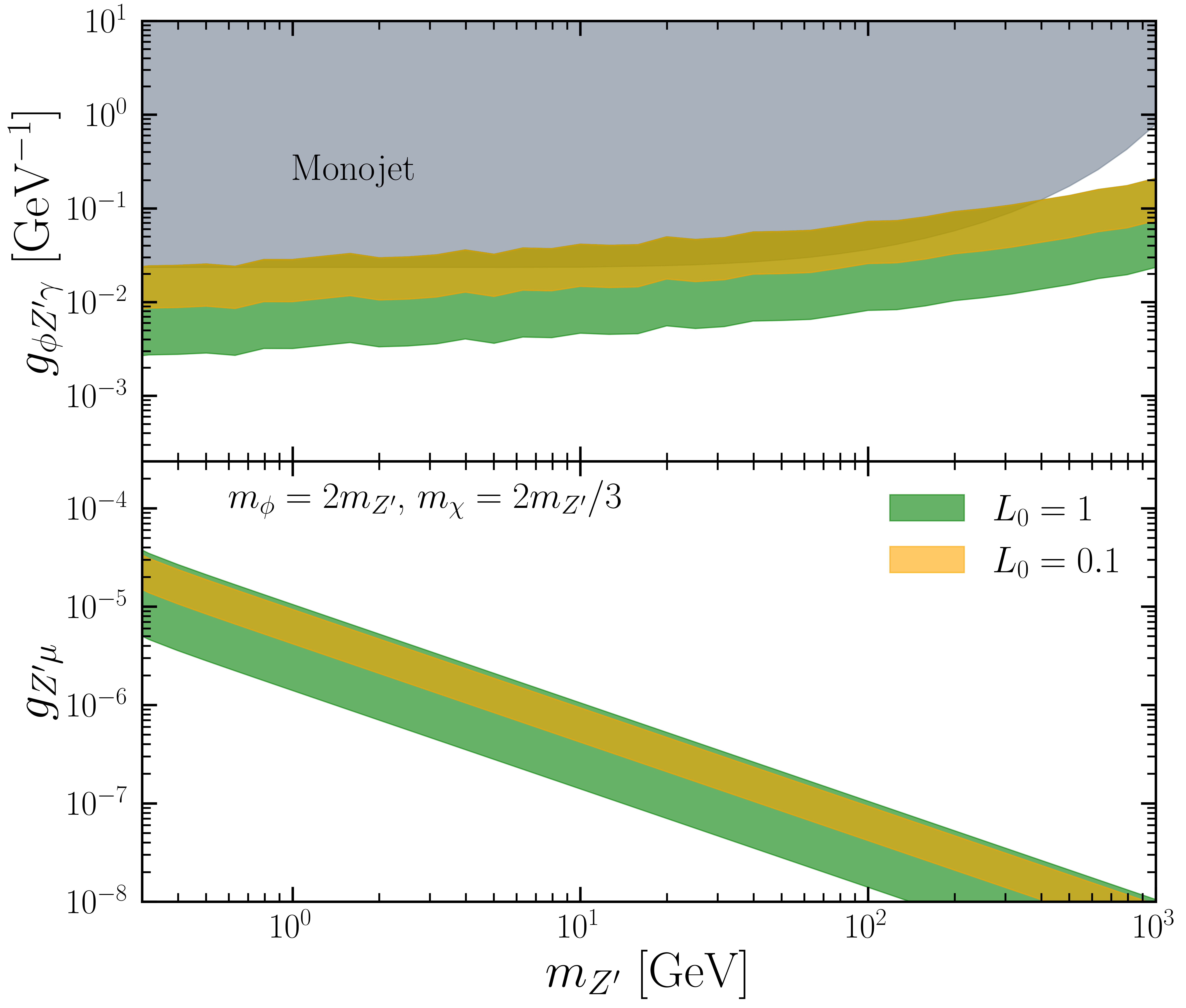}
        \caption{}
        \label{fig:coupmass_el}
    \end{subfigure}
    \captionsetup{justification=Justified}
    \caption{Preferred couplings and masses to explain the KM3NeT event for various luminosities. Existing  bounds from lab searches are shown in grey.} 
    \label{fig:coupmass}
\end{figure*}

{\textbf{Model Parameters and Constraints.--}}
In Fig.~\ref{fig:coupmass}, we show the range of masses and couplings for both inelastic and elastic DM scenarios that give rise to the cross sections and lifetimes required for one event at KM3NeT. We depict our sensitivities for $300~\text{MeV} < m_{Z'} < 1~\text{TeV}$ where the requirements for forward decay (from intermediate particle) and forward scattering (from intial DM) impose the lower and higher limit on $m_{Z'}$, respectively. In Fig.~\ref{fig:coupmass_inel}, we assume that $m_{\chi_2} = m_{\chi_1} + m_{Z'}$, and $g_{\chi} = 2.5$, and therefore show the limits for $g_{Z' q}$ and $g_{Z'\mu}$ as a function of $m_{Z'}$. The direct constraints on $g_{Z'\mu}$ come from the charged kaon~\cite{NA62:2021bji} and  pion~\cite{PIENU:2021clt} decay, as well as from $(g-2)_\mu$~\cite{Muong-2:2025xyk}. However, for heavy mediators, $m_{Z'} > 300~\text{MeV}$, these bounds appear only for $g_{Z'\mu} > 10^{-3}$. Since the lifetimes require couplings much less than $10^{-3}$, these bounds are not relevant for our study. For $g_{Z'q}$, however, we find that constraints from monophoton searches at BaBar~\cite{BaBar:2017tiz} exclude $g_{Z'q} \gtrsim 10^{-1}$ for $m_{Z'} \lesssim 10~\text{GeV}$, and the $Z$ boson decay width rule out $g_{Z'q} \gtrsim 2$ for $m_{Z'} \lesssim 80~\text{GeV}$. Here, we assume that the mixing between $Z'-\gamma$ is $\sim e/(4\pi^2)$. For heavier $Z'$, monojet searches at LHC~\cite{CMS:2021far} rule out $g_{Z'q} \gtrsim 10^{-2}$.

In the elastic DM model, we assume that $m_{\phi} = 2 m_{Z'}$, $m_{\chi} = 2m_{Z'}/3$, and $g_{\chi} = 2.5$. Under these assumptions, Fig.~\ref{fig:coupmass_el} shows the required couplings $g_{\phi Z' \gamma}$ and $g_{Z'\mu}$ as a function of $m_{Z'}$ that satisfy the required cross sections and lifetimes. Since the scalar $\phi$ decays invisibly once produced and $Z'$ is not allowed kinematically to decay into a single photon, we find that monophoton searches at DELPHI and BaBar do not apply here. However, the monojet cross section bound from ATLAS~\cite{ATLAS:2021kxv} for $p_T>200~\text{GeV}$ is translated into a constraint  on $g_{\phi Z' \gamma}$.


{\textbf{Discussion.--}}
To explain the KM3NeT event, our scenarios require an observed blazar luminosity of ${\cal O}(10^{49}~{\rm  erg/s})$. This is two orders of magnitude greater than that of TXS 0506+056, previously observed by IceCube and Fermi~\cite{IceCube:2018cha,Fermi-LAT:2019hte}. The apparent absence of such bright blazars in Fermi data may be attributed to either intergalactic magnetic field effects~\cite{Fang:2025nzg,Crnogorcevic:2025vou, Das:2025vqd} or Compton thick source environment~\cite{Hickox:2018xjf}. Producing a skymap of candidate blazar sources with the right chord length that could give rise to observable DM-induced events at IceCube in our framework is a worthwhile exercise that is left as future work. 

In both models, DM scattering within the Earth leads to highly collimated $\mu^+ \mu^-$ final states, either from the decay of a heavier DM component or a long-lived mediator. These signatures are testable at KM3NeT and IceCube and may also affect IceCube’s flavor-triangle analyses. The scenarios are subject to existing constraints from LHC monojet searches, as well as from LEP and BaBar. Notably, similar frameworks have been explored to account for the low-energy excesses observed by MiniBooNE and MicroBooNE~\cite{Dutta:2025fgz, Dutta:2021cip}. 

It is also possible to explain the ANITA-IV anomalous  events with comparable chord lengths as the KM3NeT event~\cite{ANITA:2020gmv}, within our DM model parameter spaces, without conflicting with IceCube and Pierre Auger~\cite{PierreAuger:2025hvl} non-observations, by considering different blazar sources located along the direction of the events. However, the ANITA-I and III anomalous events with steep angles~\cite{ANITA:2018sgj} are difficult to explain.  

 Additionally, a variety of DM models can be probed using blazars with known luminosities (e.g., TXS 0506+056), by analyzing the resulting leptonic and hadronic signatures produced through scattering processes at IceCube and KM3NeT. Thus, our proposal opens up a new avenue to explore blazar-boosted DM at neutrino telescopes.

\acknowledgments
{\textbf{Acknowledgments.--}} We thank Carlos Arg\"{u}elles, Sebastian B\"{o}ser, Vedran Brdar, Dibya Chattopadhyay, Saurav Das, Peter Denton,  Raj Gandhi, Gordan Krnjaic, Pedro Machado, Danny Marfatia, Nityasa Mishra, Subir Sarkar, Deepak Sathyan, Thomas Schwemberger, and Krista Smith for many useful discussions. We also thank Yasaman Farzan and Matheus Hostert for discussions and for coordinating the arXiv submission of their related work~\cite{Farzan:2025ydi}. The work of PSBD and WM was partly supported by the U.S. Department of Energy under grant No.~DE-SC0017987. PSBD was also partly supported by a Humboldt Fellowship from the Alexander von Humboldt Foundation. The work of BD, AK, LES, and AV is supported by the U.S. DOE
Grant~DE-SC0010813. 
\bibliography{main}

\clearpage
\onecolumngrid

\begin{center}
\textbf{\large Supplemental Material}
\end{center}

\setcounter{section}{0}
\renewcommand{\thesection}{\Alph{section}}
\renewcommand{\theequation}{\thesection\arabic{equation}}

\section{I. Comment on Decaying DM Solution}
\begin{figure}[!t]
    \centering
    \includegraphics[width=0.5\linewidth]{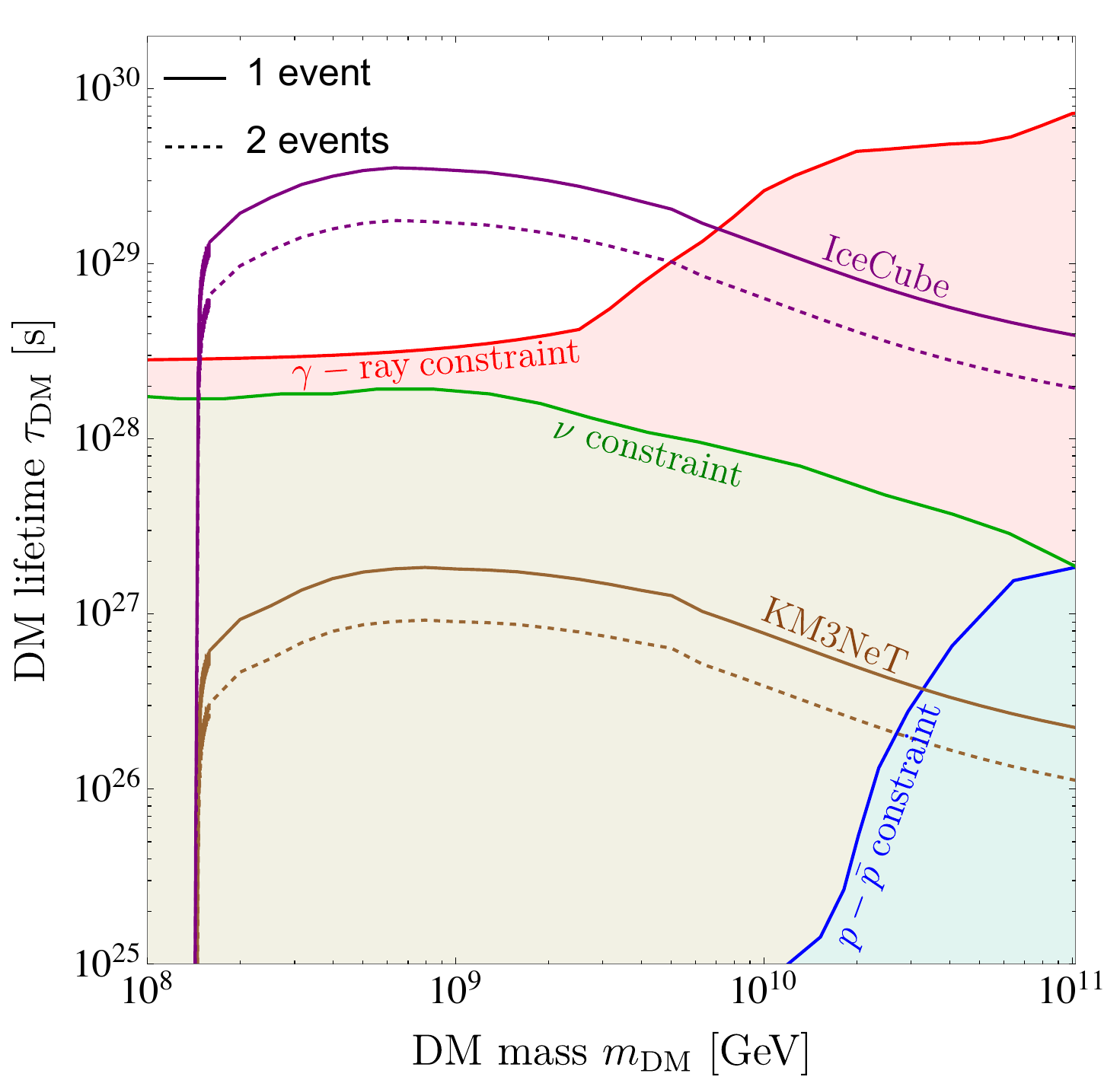}
    \captionsetup{justification=Justified}
    \caption{\rm DM lifetime as a function of its mass for KM3NeT and IceCube to detect $1$ (solid) or 2 (dashed) events  within the energy bin corresponding to the $220$ PeV KM3NeT event. The existing constraints from DM decay to neutrinos in the neutrino~\cite{Arguelles:2022nbl}, gamma-ray and $p-\bar{p}$~\cite{Das:2023wtk} channels are shown as red, green, and blue shaded regions, respectively.}
    \label{fig:EventDMdecay}
\end{figure}

Here we show explicitly why a  decaying heavy DM solution, as proposed in Refs.~\cite{Borah:2025igh, Kohri:2025bsn, Narita:2025udw,Jho:2025gaf, Barman:2025hoz,Murase:2025uwv, Khan:2025gxs}, can neither explain the KM3NeT event nor address the tension between KM3NeT and IceCube.  
In particular, we find that the neutrino flux obtained from the decay of such DM with mass ($m_{\rm DM}$) around $440$ PeV, to explain the energy of the event, and lifetime ($\tau_{\rm DM}$) at least $10^{29}$ sec, in order to satisfy the existing constraints, would give $\approx 0.01$ events in the concerned KM3NeT energy bin. In other words, 1 event at KM3NeT would require a DM lifetime two orders of magnitude smaller which is firmly excluded by both gamma-ray~\cite{Das:2023wtk} and neutrino~\cite{Arguelles:2022nbl} constraints, as shown in Fig.~\ref{fig:EventDMdecay} where we have explored the possible $m_{\rm DM}-\tau_{\rm DM}$ parameter space for KM3NeT to detect $1$ and $2$ events in the concerned energy bin. Here the DM is considered to be neutrinophilic -- the most optimistic scenario for KM3NeT. On the other hand, for the allowed parameter space, IceCube would already have seen 1 event in the same energy bin, which highlights the tension with KM3NeT for the DM decay solution.    

The differential flux of neutrinos and anti-neutrinos per neutrino flavor $\alpha$ created by the DM decays has two components, namely, galactic and extragalactic: 
\begin{equation}
\frac{d\Phi^\mathrm{\rm DM}_{\nu_\alpha+\bar{\nu}_\alpha}}{dE_\nu d\Omega}= \frac{d\Phi^\mathrm{Gal}_{\nu_\alpha+\bar{\nu}_\alpha}}{dE_\nu d\Omega}+\frac{d\Phi^\mathrm{ExtGal}_{\nu_\alpha+\bar{\nu}_\alpha}}{dE_\nu d\Omega} \, .
\end{equation}
The galactic component has the following form:
\begin{align}
\frac{d\Phi^{\mathrm{Gal}}_{\nu_\alpha+\bar{\nu}_\alpha}}{dE_\nu}=&\frac{1}{4\pi m_{\rm DM}\tau_{\rm DM}}\frac{dN_\alpha}{dE_\nu}\int d\Omega (l,b) 
\int_0^{\infty} ds\rho_{\rm DM}(r(l,s,b)) \, .
\end{align}
Note that the radial distance $r$ depends on the galactic coordinates $l$ and $b$, and the line-of-sight distance $s$ from the Earth, i.e. $r=\sqrt{s^2+R_\odot^2-2sR_\odot \cos l \cos b}$ with $R_\odot=8.5$~kpc for Milky Way. We assume that the galactic DM density distribution $\rho_\mathrm{\rm DM}(r)$ follows the NFW profile~\cite{Navarro:1996gj}:
\begin{equation}
    \rho_{\rm DM}(r)=\frac{\rho_s}{\frac{r}{r_s}\left(1+\frac{r}{r_s}\right)^2} \, ,
\end{equation}
with $r_s = 24$ kpc and $\rho_s = 0.18$ GeV cm$^{-3}$ for Milky Way~\cite{Cirelli:2010xx}.

The extragalactic component is given by
\begin{align}
\frac{d\Phi^{\rm Ext. Gal}_{\nu_\alpha+\bar{\nu}_\alpha}}{dE_{\nu} d\Omega}=\frac{\rho_{\rm DM}}{4\pi m_{\rm DM}\tau_{\rm DM}}\int_0^\infty (1+z)\frac{dz}{H(z)}\frac{dN_\alpha}{dE_\nu} \Big |_{E_\nu(1+z)} \, ,
\end{align} 
where $\rho_{\rm DM}=\Omega_{\rm DM}\rho_c$, with $\rho_c=5.5\times 10^{-6}$ GeV cm$^{-3}$ being the critical density of the Universe  and $\Omega_{\rm DM}=0.268$ the relic abundance of DM , $z$ is the cosmological redshift and $H(z)=H_0\sqrt{\Omega_\Lambda + \Omega_M(1+z)^3}$ is the Hubble expansion rate, with the dark energy and the matter cosmic energy densities $\Omega_\Lambda=0.685$, $\Omega_M=0.315$, and the Hubble constant $H_0 = 67.3$~km~s$^{-1}$~Mpc$^{-1}$ ~\cite{Planck:2018vyg}.

The differential flux of neutrinos and anti-neutrinos in both galactic and extragalactic components depends on the energy spectrum $\frac{dN_\alpha}{dE_\nu}$ of the  $\alpha$-flavored neutrinos generated by the decay of a single DM particle. We obtain these energy spectra using the  \texttt{HDMSpectra}~\cite{Bauer:2020jay} code.

The number of events at KM3NeT or IceCube can then be calculated as
\begin{equation}
\label{eq:Nw}
N_\mathrm{event} = T \Omega\int_{E_\nu^\mathrm{min}}^{E_\nu^\mathrm{max}} dE_\nu  A_\mathrm{eff}(E_\nu) \frac{d\Phi^{\mathrm{\rm DM}}_{\nu_\alpha+\bar{\nu}_\alpha}}{dE_\nu}(E_\nu) \, , 
\end{equation}
where $T$ is the exposure time ($335$ days for KM3NeT and 3577 days for IceCube), $\Omega=4\pi$ for diffuse all-sky average flux, $E_{\nu}^\mathrm{min} = 10^{7.9}$ GeV and $E_{\nu}^\mathrm{max}=10^{9.4}$ GeV for the energy bin as in the KM3NeT analysis~\cite{KM3NeT:2025npi}, and $A_{\mathrm{eff}}$ is the effective area which is a function of energy~\cite{KM3NeT:2025npi,  IceCube:2014vjc}. 

\section{II. DM Production in Blazar and Source Spectra}
In this section we compare the expected DM and neutrino source spectra produced from a blazar and explain why the DM flux overtakes the neutrino flux at the highest energies.
To simulate the neutrino/DM flux, we model an internal‐shock region moving with bulk Lorentz factor $\Gamma\sim10$. 
We assume the apparent bolometric radiation luminosity of the jet $L_{\mathrm{rad}}\sim10^{48}$erg/s. 
Considering a spherical, relativistic ``blob", its comoving size is $l_b \approx \Gamma c \delta t^{\prime}$, where $\delta t^{\prime}$ is the variability time in black hole frame which we take to be $10^5$~s.
The dynamical time scale of the system is $t_{\mathrm{dyn}} \approx l_b / c$.
The shock-accelerated proton spectrum is modeled as a power‑law with exponential cutoff: 
\begin{equation}
   \frac{ d N_p}{dE_p}\left(E_p\right)=A_0 E_p^{-\alpha_p} \exp \left(-E_p / E_{\text {cut }}\right), \quad \alpha_p=2.2, \, E_{\text {cut }}=10^8 ~\mathrm{GeV} \, .
\end{equation}

High‑energy protons accelerated in the jet interact both with the jet’s internal radiation field and with any thermal photon population, initiating photohadronic ($p\gamma$) processes. 
In the jet comoving frame, the energy density of the non‑thermal photon field produced inside the blazar jet can be estimated as
\begin{equation}
U_\gamma \approx \frac{3 L_{\mathrm{\gamma}}}{4 \pi \Gamma^4 l_b^2 c}\approx 1~\mathrm{erg}/\mathrm{cm}^3 \,. 
\end{equation}
Furthermore, relativistic protons can interact with thermal photon populations supplied by external structures such as the accretion disc, the broad‑line region (BLR), and the dust torus. For our estimates we use the total target photon density in the comoving frame from Refs.~\cite{Murase:2014foa, Kalashev:2022scs}.

In addition, these protons can undergo hadronic ($pp$) collisions with cold, quasi‑stationary protons in the surrounding medium.
Assuming proton kinetic power $L_p\sim10^{49}$erg/s the comoving cold proton density in the jet is approximately given by
\begin{equation}
n_p \approx \frac{3 L_p}{4 \pi \Gamma^4 l_b^2 m_pc^3}\approx5.89\times10^3/\mathrm{cm}^3 \, .
\end{equation}
The numerical values adopted for the comoving photon number density $n_\gamma$ and cold proton density $n_p$ serve only as illustrative benchmarks. In practice, both quantities are dynamic and depend on several engine and environment specific factors.

\subsection*{Neutrino flux}

Using the photopion production cross-section $\sigma_{p \gamma}$ and inelasticity $\kappa_{p \gamma}$ ~\cite{2009herb.book.....D},
the photomeson energy loss time scale is given by \cite{Oikonomou:2019djc}
\begin{equation}
t_{p \gamma}^{-1}\left(E_p\right)=\frac{c}{2 \gamma_p^2} \int_{\varepsilon_{r, \min }}^{\infty} \mathrm{d} \varepsilon_r \sigma_{p \gamma}\left(\varepsilon_r\right) \kappa_{p \gamma}\left(\varepsilon_r\right) \varepsilon_r \int_{\varepsilon_r /\left(2 \gamma_p\right)}^{\infty} \frac{n_\gamma\left(\varepsilon\right)}{\varepsilon^{ 2}} \mathrm{~d} \varepsilon \, ,
\end{equation}
where $n_\gamma$ is the target photon number density in the comoving frame; $ \gamma_p$ is the proton Lorentz factor in the comoving frame,  $\varepsilon_r$ is the photon energy in the rest frame of proton and $\varepsilon_{r,{\rm min}}\sim145~\text{MeV}$ the threshold photon energy for photomeson production.
Similarily $t_{pp}^{-1}=n_p\sigma_{p p} \kappa_{p p} c$ for the $pp$ production channel.
The dimensionless efficiency entering the meson (and hence neutrino) production is
\begin{equation}
f_{p \gamma/pp}\left(E_p\right) \approx \frac{t_{\mathrm{dyn}}}{t_{p \gamma/pp}} \,.
\end{equation}

The comoving neutrino production spectrum produced by pion decay and subsequent muon decay therefore follows from 
\begin{equation}
E_\nu^{ 2} \frac{\mathrm{~d} N_\nu^{}}{\mathrm{d} E_\nu^{}}\approx\frac{3}{8} f_{p \gamma / p p} f_{\pi,\text{cool}} E_p^{ 2} \frac{ d N_p}{dE_p} \, ,
\end{equation}
where the produced neutrinos carry only a small fraction of the parent proton energy $E_\nu \simeq 1/20 E_p$; $f_{\pi, \text { cool }}=1-\exp \left(-t_{\pi, \mathrm{cool}} / t_{\pi, \mathrm{dec}}\right)$.
The pion cooling time $t_{\pi,\text{cool}}$ is determined by combining the inverse of the synchrotron cooling time with the inverse of the dynamical timescale $t_{\pi,\text{cool}}^{-1}=t_{\pi, \text { syn }}^{-1}+t_{\mathrm{dyn}}^{-1}$  \cite{Oikonomou:2019djc}.
For neutrino production, the $pp$ interactions are expected  to be neglegible compared to $p\gamma$ production.

\subsection*{Dark Matter flux}
For the DM production channels through $pp$ and $p\gamma$ interactions, the resulting flux is
\begin{equation}
E_\chi^{ 2} \frac{\mathrm{~d} N_\chi}{\mathrm{d} E_\chi^{}}= x f_{\text {int }} E_p^{ 2} \frac{ d N_p}{dE_p} \, ,
\end{equation}
where $x=0.5$ is the approximate fraction of the parent proton energy carried by the DM particle. Here the DM production efficiency  $f_{\text {int}}$ is defined similar to the case of neutrinos for both $pp$ and $p\gamma$ channels with the corresponding neutrino production cross-sections and inelasticities replaced by the DM counterpart.

\begin{figure}[t!]
    \centering
    \includegraphics[width=0.5\linewidth]{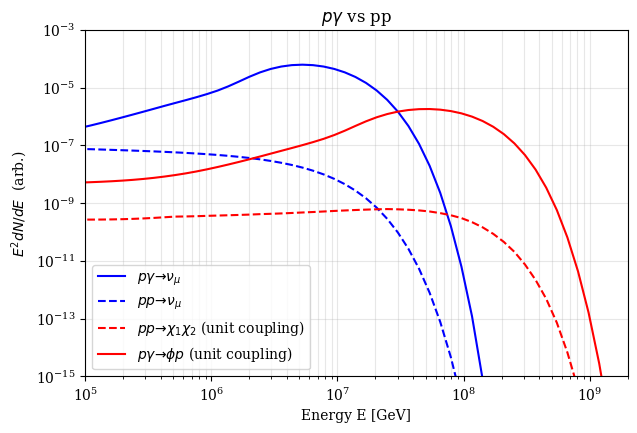}
\captionsetup{justification=Justified}    \caption{Comoving‐frame expected DM and neutrino spectra $E^2 \mathrm{~d} N / \mathrm{d} E$ in arbitrary units. The DM channels overtake the neutrino channels above $E\sim10^8$ GeV.}
    \label{fig:prodSpectra}
\end{figure}

Figure~\ref{fig:prodSpectra} shows the comoving $E^2dN/dE$ spectra for neutrinos and DM
obtained with the benchmark parameters above. At energies $E>100$~PeV, the DM flux from either $p\gamma$ or $pp$ production could significantly exceed the accompanying neutrino flux.  This is because the assumed DM production channels deposit a larger fraction of the proton energy, than photopion neutrino production, amplifying the high‑energy yield. In addition, unlike charged pions and muons there are no radiative losses since the neutral DM pair does not suffer synchrotron cooling. So its spectrum inherits the full high‑energy power‑law tail of the parent protons. This justifies why the DM explanation proposed here is only relevant at the highest energies, while the ``lower" energy events observed by IceCube can still be the  neutrino-induced events.

\section{III. Effective Area Calculation}
Here we calculate the general effective area for DM single and multiple scattering scenarios, where the initial state $\chi_i$ produces an intermediate state $X$ which then produces $\mu^+ \mu^-$. We define $\gamma_s = 1/\lambda_{\rm{scatter}}$, and $\gamma_d = 1/\lambda_{\rm{decay}}$, the $\lambda$'s being the corresponding MFPLs in Earth matter. Given a muon with initial energy $E_\mu$, we define $d_{\text{th}}(E_\mu)$ as the distance it propagates such that its final energy is $E_{\text{th}}$. In other words, this is the maximum threshold distance outside the detector such that the muon can still be detected above certain threshold energy $E_{\text{th}}$. For a muon with energy $E_\mu$ produced at a distance $d_y = d_b-x-y$ outside the detector, where $d_b$ is the total Earth overburden, $x$ and $y$ are the distances traveled by $\chi_i$ and $X$, respectively, we define $E_{\mu}^f(d_y)$ to be the energy of the muon after it travels a distance $d_y$. Both $E_{\rm th}$ and $E_{\mu}^f(d_y)$ can be calculated using the energy loss equation of muons in a medium~\cite{Gaisser:2016uoy}. Therefore, the effective area for single and multiple scattering scenarios can be defined as follows:
    \begin{equation}
        \begin{aligned}
            A_{\text{eff}}^{\text{single}}(N_\text{PMT},E_{\chi_i},E_{\text{th}}, d_{\rm{b}}, \delta) =& A_{\text{csec}}\int dE_{X}P(E_{X}|E_{\chi_i})\int dE_{\mu} P(E_\mu|E_{X})\int^ {d_{\rm{b}} - d_{\text{th}}(E_\mu)}_{0} dx \gamma_s e^{-\gamma_s x} \\& \times \bigg[ \int_{d_{\rm{b}} - d_\text{th}(E_\mu) - x}^{d_{\rm{b}} - x} dy~P_{N_{\text{PMT}}}(E_{\mu}^f(d_y))\gamma_de^{-\gamma_d y} + \int_{d_{\rm{b}} - x}^{d_{\rm{b}} + \delta - x} dy~P_{N_{\text{PMT}}}(E_{\mu}) \gamma_d e^{-\gamma_dy} \bigg]\, ,
        \end{aligned}        \label{eq:DMEffarea_single}
    \end{equation}
    \begin{equation}
        \begin{aligned}
            A_{\text{eff}}^{\text{multiple}}(N_\text{PMT},E_{\chi_i},E_{\text{th}}, d_{\rm{b}}, \delta) = &A_{\text{csec}} \int dE_{X}P(E_{X}|E_{\chi_i})\int dE_{\mu} P(E_\mu|E_{X}) \gamma_d\bigg[ \int_{d_{\rm{b}} - d_\text{th}}^{d_{\rm{b}}} dx~P_{N_{\text{PMT}}}(E_{\mu}^f(d_y)) P_{X}(x)
            \\& \qquad + \int_{d_{\rm{b}}}^{d_{\rm{b}} + \delta} dx~P_{N_{\text{PMT}}}(E_{\mu}) P_{X}(x) \bigg] \, ,
        \end{aligned}   
        \label{eq:DMEffarea_multiple}
    \end{equation}
where $A_{\rm csec}$ is the geometric cross sectional area of the detector along the event direction, $\delta$ is the detector dimension along the direction of propagation, $N_{\rm PMT}$ is the number of PMTs triggered, and $P(E_a|E_{\chi_i})$ and $P(E_\mu|E_{X})$ are defined as follows:
\begin{equation}
    P(E_X|E_{\chi_i}) = \frac{1}{\sigma_{\chi_i\to X}(E_{\chi_i})}\frac{d\sigma_{\chi_i\to X}(E_{\chi_i})}{dE_{X}} \, , \quad P(E_\mu|E_{X}) = \frac{1}{\Gamma_{X\to \mu}(E_{X})}\frac{d\Gamma_{X\to \mu}(E_{X})}{dE_{\mu}} \,.
\end{equation}

The effective area formulated for multiple scattering is valid under two conditions: (i) The scattering and re-scattering processes have the same probability, i.e., $P(\chi_i \to X) = P(X \to \chi_i$), and (ii) majority of the initial-state-energy is transferred to the final state in each scattering, i.e., $P(E_X|E_{\chi_i}) = P(E_{\chi_i}|E_{X}) = \delta(E_X-E_{\chi_i})$. Under these two conditions,
\begin{equation}
    P_{X}(x) = \gamma_d\gamma_se^{-\big( \gamma_d +2\gamma_s \big)x/2} \frac{\sinh \big( x/2\sqrt{\gamma_d^2 + 4\gamma_s^2} \big)}{1/2\sqrt{\gamma_d^2 + 4\gamma_s^2}} \, .
\end{equation}


In the inelastic DM scenario, we realize that the above conditions are satisfied when $m_{Z'},m_{\chi_{1,2}} \lesssim 100$~GeV. We also find that $P(E_\mu|E_{\chi_2}) = \delta(E_\mu-E_{\chi_2})$ as long as $m_{\chi_2} \simeq m_{\chi_1} + m_{Z'}$ and $m_{\chi_1} \leq 2m_{\mu}$. For the elastic DM scenario,  the processes $\chi_i \to X$ and $X \to \chi_1$  correspond to $\chi+N \to \chi + N+ Z'$ ($X = Z'$) and $Z' + N \to \phi (\to \chi \bar{\chi}) + N$, respectively. In the former process, $Z'$ typically carries 80\% of the incoming energy of DM, with a probability of 30\%. As a result, the re-scattering results in only $\sim 50\%$ transfer of energy from the $Z'$ to each $\chi$. Since DM loses $\sim 40\%$ of its initial energy in each scatter + re-scatter step, multiple scattering is not as efficient as in the inelastic scenario. Therefore, we estimate the event rates given in the main text using multiple scattering for the inelastic DM scenario and single scattering for the elastic case.

\begin{figure}[!t]
    \centering
    \includegraphics[width=0.5\linewidth]{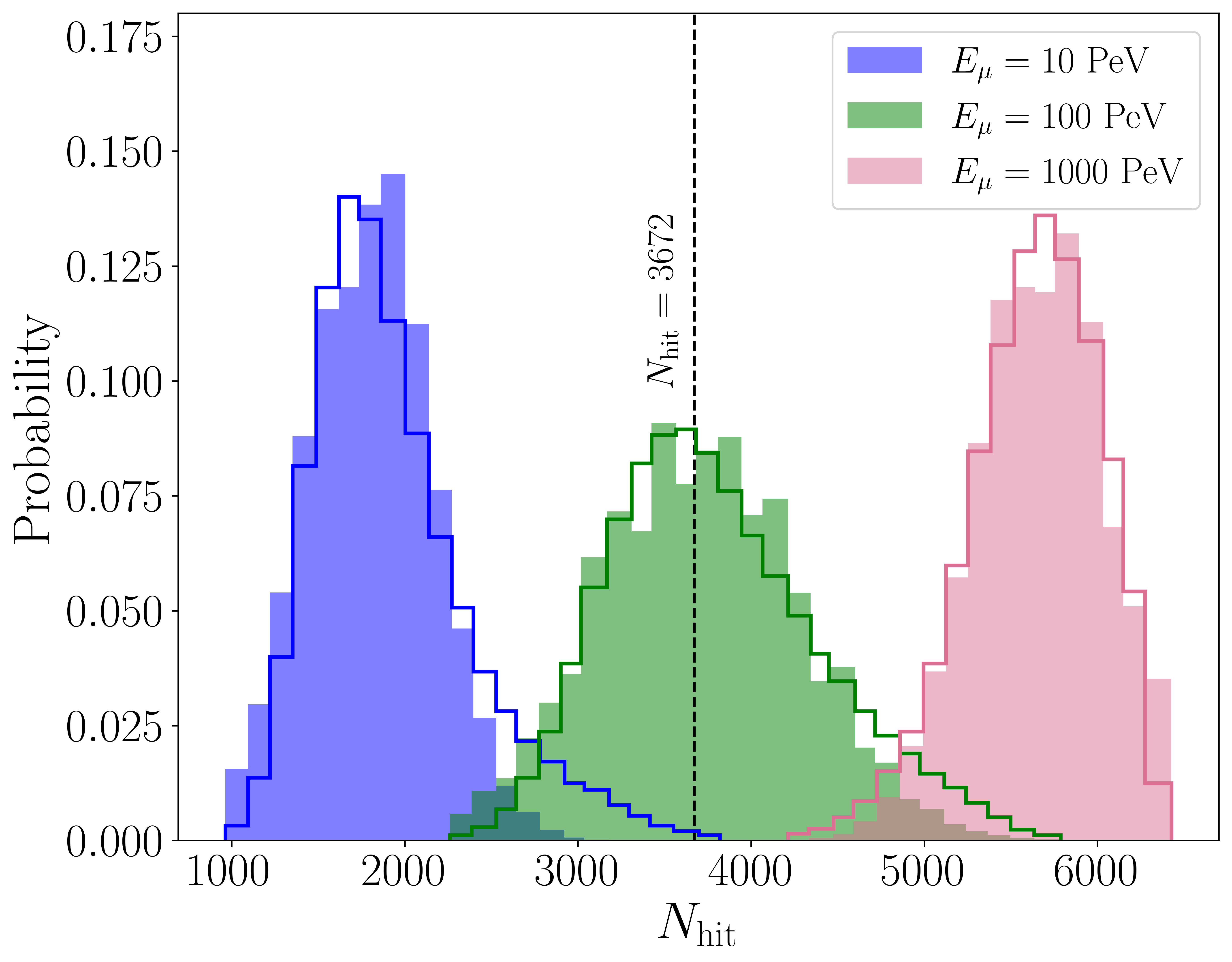}
    \captionsetup{justification=Justified}
    \caption{ The Gaussian distributions (shaded histograms) derived for the number of PMTs triggered for 10, 100, and 1000 PeV muons alongside the official results (solid lines)~\cite{KM3NeT:2025npi}. The dashed vertical line labeled ``3,672 PMTs" corresponds to that observed for the KM3-230213A event.}
    \label{fig:PNhit}
\end{figure}

The prior $P_{N_{\text{PMT}}} (E_{\mu})$ requires that the muon with energy $E_\mu$ triggers atleast $N_{\text{PMT}}$ number of PMTs at the detector. For the event observed at KM3NeT, this implies at least 3672~PMTs. We utilize the mapping between the muon energy and the probability distribution of $N_{\text{PMT}}$ as shown in Ref.~\cite{KM3NeT:2025npi} and fit them to a Gaussian distribution. We then interpolate between the three available means and the variances of these distributions to find the distributions for any given muon energy, i.e., $\mu(E_\mu), \sigma^2(E_{\mu})$. Our results are shown in the shaded histograms in Fig.~\ref{fig:PNhit}, where we see that the approximated Gaussian fits are close to the official results~\cite{KM3NeT:2025npi}, which are shown by the solid lines. Based on the Gaussian approximations, the prior on the required PMT hits is 
\begin{equation}
    P_{N_{\text{PMT}}} (E_{\mu}) = \frac{1}{2} \bigg( 1 - \text{erf}\bigg( \frac{N_{\text{PMT}} - \mu(E_\mu)}{\sqrt{2}\sigma(E_{\mu})}\bigg)\bigg) \, .
\end{equation}

By comparing the number of events using Gaussian fits for the diffuse neutrino flux predictions in Ref.~\cite{Li:2025tqf}, we find that the Gaussian fits to the probability distributions are close to those found using the Fretchet distributions.

\end{document}